\documentclass[twocolumn]{aastex631}
\usepackage[utf8]{inputenc}
\usepackage{amsmath}
\usepackage{amssymb}
\usepackage{graphicx}
\usepackage{xcolor}
\usepackage{enumitem}

\received{---}
\revised{---}
\accepted{---}

\submitjournal{ApJ}

\shorttitle{Implications of Equilibrium States}
\shortauthors{Voit et al.}

\begin{document}

\title{\textbf{Equilibrium States of Galactic Atmospheres II: Interpretation and Implications}}

%\correspondingauthor{G. M. Voit}

\author[0000-0002-3514-0383]{G. Mark Voit}
\affiliation{Michigan State University,
Department of Physics and Astronomy,
East Lansing, MI 48824, USA}

\author[0000-0002-5840-0424]{Christopher Carr}
\affiliation{Columbia University,
Department of Astronomy,
New York, NY 10027, USA}

\author[0000-0003-3806-8548]{Drummond B. Fielding}
\affiliation{Center for Computational Astrophysics, 
Flatiron Institute, 
New York, NY 10010, USA}
\affiliation{Department of Astronomy, Cornell University, Ithaca, NY 14853, USA}

\author[0000-0002-2499-9205]{Viraj Pandya}
\altaffiliation{Hubble Fellow}
\affiliation{Columbia University,
Department of Astronomy,
New York, NY 10027, USA}

\author[0000-0003-2630-9228]{Greg L. Bryan}
\affiliation{Columbia University,
Department of Astronomy,
New York, NY 10027, USA}

\author[0000-0002-2808-0853]{Megan Donahue}
\affiliation{Michigan State University,
Department of Physics and Astronomy,
East Lansing, MI 48824, USA}

\author[0000-0003-4754-6863]{Benjamin D. Oppenheimer}
\affiliation{University of Colorado, 
Center for Astrophysics and Space Astronomy, 389 UCB, 
Boulder, CO 80309, USA}

\author[0000-0002-6748-6821]{Rachel S. Somerville}
\affiliation{Center for Computational Astrophysics,
Flatiron Institute,
New York, NY 10010, USA}

%\correspondingauthor{G. M. Voit}
%\email{voit@msu.edu}

\begin{abstract}
% \begin{center} 
%    \quad \quad \quad \quad \marknotes{Abstract goes here}
% \end{center}
The scaling of galaxy properties with halo mass suggests that feedback loops regulate star formation, but there is no consensus yet about how those feedback loops work. To help clarify discussions of galaxy-scale feedback, Paper I presented a very simple model for supernova feedback that it called the \textit{minimalist regulator model}. This followup paper interprets that model and discusses its implications. The model itself is an accounting system that tracks \textit{all} of the mass and energy associated with a halo's circumgalactic baryons---the central galaxy's \textit{atmosphere}. Algebraic solutions for the equilibrium states of that model reveal that star formation in low-mass halos self-regulates primarily by expanding the atmospheres of those halos, ultimately resulting in stellar masses that are insensitive to the mass-loading properties of galactic winds. What matters most is the proportion of supernova \textit{energy} that couples with circumgalactic gas. However, supernova feedback alone fails to expand galactic atmospheres in higher-mass halos. According to the minimalist regulator model, an atmospheric contraction crisis ensues, which may be what triggers strong black-hole feedback. The model also predicts that circumgalactic medium properties emerging from cosmological simulations should depend largely on the \textit{specific energy} of the outflows they produce, and we interpret the qualitative properties of several numerical simulations in light of that prediction.
\end{abstract}

\section{Introduction}

Each galaxy is a vast and complex dynamical system with innumerable internal degrees of freedom, driven by a broad array of external environmental influences. Yet galaxies as a population conform to some relatively simple scaling relations connecting a galaxy's stellar mass and star formation rate with the mass of the cosmological halo surrounding it \citep[e.g.,][]{Moster_2010ApJ...710..903M,Behroozi+2013ApJ...770...57B,Behroozi2019MNRAS.488.3143B}. If you ask an astronomer in a hurry how those scaling relations arise, you might get a one-word reply: feedback. But that's not a satisfactory answer. Multitudes of hypotheses, models, and numerical simulations have been devoted to unpacking what that single word means, without yielding a clear consensus about how galaxy-scale feedback processes work \citep[e.g.,][]{SomervilleDave_2015ARA&A..53...51S,NaabOstriker2017ARA&A..55...59N,Crain_vandeVoort_2023ARA&A..61..473C}.

Astronomers generally agree about many aspects of the overall story: Galaxies grow inside potential wells with origins dating back to the universe's earliest moments. Gravity causes those potential wells to deepen over time. Most of that gravitating matter is thought to be non-baryonic, but about 16\% of the mass consists of baryons. The baryons can radiate away some of their potential energy and gather into the collections of stars we call galaxies, but stars release considerable amounts of energy, especially when they start to explode, and how those explosions affect galaxy evolution is not yet clear. 

Supernova energy can limit star formation in more than one way. It might eject some of a galaxy's gas before it can form stars---we will call this feedback mode \textit{ejective}. Or it might reduce a galaxy's gas supply by preventing circumgalactic gas from entering it---we will call this feedback mode \textit{preventative}. Meanwhile, supernova-driven interstellar turbulence might \textit{delay} conversion of a galaxy's gas into stars.\footnote{This paper does not explicitly consider interstellar turbulence because the equilibrium states of galactic atmospheres do not strongly depend on it, as long as a galaxy's star formation timescale is short compared with the time required for the galaxy's atmosphere to approach a feedback-regulated equilibrium state (see \S 3.9 of Paper I).} One of those modes might be more important than the others, but modeling all three might be necessary. Furthermore, exploding stars might not be the only feedback source, because gas accreting onto the supermassive black holes at the centers of galaxies can potentially release far more energy. Attempts to model galaxy-scale feedback must therefore make choices about how to implement feedback from both supernovae and black-hole accretion and also about how the energy released interacts with a galaxy's environment.

This paper and its companion (Voit et al. 2024, hereafter Paper I) look at those choices from a birds-eye perspective, ignoring many details in order to clarify the consequences of the choices. Our approach emphasizes the equilibrium states of the feedback-regulated systems resulting from various implementations of supernova feedback.\footnote{While Paper I focuses on supernova feedback, future papers in this series will show how the overall approach can be extended to include energy released during black-hole accretion.} Emphasizing equilibrium might seem odd, given the chaotic morphology of feedback outbursts in numerical simulations of galaxy evolution, not to mention the disruptive merger events that occasionally accompany cosmological structure formation. With all that going on, why bother exploring a galaxy's equilibrium states? 

For a detailed answer to that question, see Paper I. It presents a highly simplified model for feedback interplay between a galaxy and its circumgalactic medium (CGM), which we will sometimes call the galaxy's \textit{atmosphere}. By expressing that interplay in terms of just three ordinary differential equations, Paper I was able to derive algebraic expressions for the equilibrium states of the simplified system, illustrating how supernova feedback self-regulates in response to atmospheric expansion and contraction. A galaxy might never settle into one of those equilibrium states, but it is continually converging toward one of them during the time intervals between feedback outbursts and merger events. Consequently, sets of those equilibrium states define \textit{asymptotic scaling relations} toward which galaxy populations converge.

This followup paper focuses on \textit{interpretation} of those equilibrium states and their \textit{implications} for both numerical simulations and semi-analytic models of galaxy evolution. We are providing that discussion here, in a separate paper, for readers more interested in Paper I's implications than its details. If you have not yet read Paper I, we recommend paying close attention to section \ref{sec:Equilibrium} of this paper, which recaps Paper I's key results and considers how observations can be used to assess whether supernova feedback is primarily ejective or preventative. 

Section \ref{sec:Predictions} then discusses how Paper I's models are related to prior semi-analytic models of galaxy evolution. Within that landscape of models, Paper I's models are most similar to the ones of \citet{Carr_2023ApJ...949...21C}, who found that star formation was insensitive to the mass-loading parameter $\eta_M \equiv \dot{M}_{\rm wind} / \dot{M}_*$ relating a galaxy's gas-mass outflow rate ($\dot{M}_{\rm wind}$) to its star formation rate ($\dot{M}_*$). Using the concept of an equilibrium state, Paper I explained \textit{why} certain models of galaxy evolution are insensitive to $\eta_M$. Section 3 of this paper expands upon that explanation and compares Paper I's models with observations of the relationship between stellar mass and halo mass.

Section \ref{sec:Simulations} shows how insights from Paper I help to explain some well known differences among cosmological simulations. Such simulations are usually optimized so that they produce populations of galaxies with star-formation histories similar to observed galaxy populations \citep[e.g.,][]{CrainEAGLE+2015,Pillepich_TNG50_2018MNRAS.475..648P}. However, numerical simulations that produce similar galaxy populations can display quite different circumgalactic medium (CGM) characteristics \citep[e.g.,][]{Crain_vandeVoort_2023ARA&A..61..473C,Wright2024}. For example, CGM masses within the virial radii of star-forming galaxies in the EAGLE simulations \citep{Schaye_EAGLE_2015MNRAS.446..521S} are less than half of what's found in the IllustrisTNG simulations \citep{Nelson_2018_bimodality,Pillepich_TNG50_2018MNRAS.475..648P}. Also, the spatial distributions of baryons beyond the virial radii of halos in the IllustrisTNG, EAGLE, and SIMBA simulations are strikingly different \citep{Sorini_2022MNRAS.516..883S,Ayromlou_2023MNRAS.524.5391A}. To quantify the extent of the baryon distributions, \citet{Ayromlou_2023MNRAS.524.5391A} defined a halo's ``closure radius" to be the smallest radius within with the baryonic mass fraction reaches the cosmic mean value $f_{\rm b}$ and found that the three simulation suites produce closure radii that depend quite differently on halo mass. As \S \ref{sec:Simulations} will show, those differences may be reflecting the \textit{specific energy} of the galactic outflows that the feedback algorithms in those simulations generate.

Section \ref{sec:Summary} summarizes the paper's findings.

%\marktext{An Appendix applies the paper's methods to some additional atmosphere models}

\newpage

\section{Equilibrium States}
\label{sec:Equilibrium}

The framework for galaxy evolution modeling presented in Paper I focuses on how imbalances between heating and cooling of a galaxy's atmosphere drive either expansion or contraction of the atmosphere. Excess heating increases the atmosphere's specific energy, leading to \textit{expansion} that lowers the atmosphere's mean gas density, lengthens both its cooling time and mean dynamical time, and reduces the galaxy's gas supply. Excess cooling decreases the atmosphere's specific energy, leading to \textit{contraction} that increases the atmosphere's mean gas density, shortens both its cooling time and mean dynamical time, and increases the galaxy's gas supply.

To capture how atmospheric expansion and contraction enable self-regulation, Paper I represents a galaxy's overall feedback system in terms of just three ordinary differential equations, giving the derivatives with respect to time of total CGM gas mass ($M_{\rm CGM}$), total CGM energy ($E_{\rm CGM}$), and interstellar gas mass ($M_{\rm ISM}$). Paper I calls that system of equations the \textit{minimalist regulator model} for galactic atmospheres. It also adopts a comprehensive accounting approach in which the fraction of a halo's baryons belonging to the CGM is 
\begin{equation}
    f_{\rm CGM} = \frac {f_{\rm b} M_{\rm halo} - M_* - M_{\rm ISM}} 
                        {f_{\rm b} M_{\rm halo}}
                        \; \; .
\end{equation}
Here, $M_{\rm halo}$ is the halo's total mass, $M_*$ is its central galaxy's stellar mass, $f_{\rm b}$ is the cosmic baryon mass fraction, and satellite galaxies are considered negligible. Baryons that feedback has pushed beyond the halo's virial radius are therefore still considered CGM baryons, and their energy contributes to $E_{\rm CGM}$. Importantly, $E_{\rm CGM}$ includes \textit{gravitational potential energy} along with thermal and non-thermal energy, meaning that $E_{\rm CGM}$ remains constant while atmospheric expansion converts both thermal and non-thermal energy into gravitational potential energy.

The minimalist regulator model has equilibrium states in which both $f_{\rm CGM}$ and the CGM specific energy
\begin{equation}
    \varepsilon_{\rm CGM} \equiv \frac {E_{\rm CGM}} {M_{\rm CGM}}
\end{equation}
remain constant. The rest of this section describes those equilibrium states and what governs them.

\subsection{Expansion and Contraction}

Whether or not a galactic atmosphere ultimately expands or contracts depends on how $\varepsilon_{\rm CGM}$ compares with the mass weighted value of $\varepsilon_{\rm acc}$, averaged over a halo's cosmic mass accretion history. If radiative cooling, galaxy formation, and feedback had never happened, then $\varepsilon_{\rm CGM}$ would be nearly equal to that average value, which is assigned to a halo's baryons early in time through cosmological initial conditions. A galactic atmosphere's current value of $\varepsilon_{\rm CGM}$ therefore reflects how radiative cooling, galaxy formation, and energy input from galactic feedback alter the initial specific energy of the halo's baryons. 

If cumulative feedback energy input exceeds cumulative radiative cooling, then $\varepsilon_{\rm CGM}$ exceeds $\varepsilon_{\rm acc}$, and the atmosphere's equilibrium state is \textit{expanded} relative to what its purely cosmological state would have been (i.e. without any radiative cooling or feedback). If instead cumulative radiative cooling exceeds cumulative feedback energy input, then the atmosphere's equilibrium state is \textit{contracted} relative to its purely cosmological state.

This paper's discussion of atmospheric expansion and contraction assumes $\varepsilon_{\rm acc} \approx 4 v_{\rm c}^2$, where $v_{\rm c}$ is the maximum circular velocity of the halo's gravitational potential well.\footnote{Paper I chooses the zero point of the halo's gravitational potential to be at its central galaxy's outer boundary, so that baryons have zero potential energy when going from the CGM into the ISM. Because of that choice, $\varepsilon_{\rm acc}$ is similar to the gravitational potential at the halo's virial radius.} The factor of 4 reflects the typical depth of a Navarro-Frenk-White potential well \citep{nfw97}. Getting out to the halo's virial radius requires $\approx 3.5 v_{\rm c}^2$, depending on how concentrated the halo's mass is toward its center, and that is the specific potential energy that accreting matter has at the halo's virial radius. The specific kinetic energy associated with infall to the virial radius is $\approx 0.5 v_{\rm c}^2$, and the sum is $\varepsilon_{\rm acc} \approx 4 v_{\rm c}^2$.

\subsection{Coupled and Uncoupled Outflows}

The minimalist regulator model of Paper I assumes that outflows of feedback energy are coupled with the CGM. For example, star formation at the rate $\dot{M}_*$ drives an outflow that transports ISM gas into the CGM at the rate
\begin{equation}
    \dot{M}_{\rm wind} = \eta_M \dot{M}_*
    \; \; ,
\end{equation}
where $\eta_M$ is the model's \textit{mass-loading parameter.} The outflow then adds energy to the CGM at the rate
\begin{equation}
    \dot{E}_{\rm fb} = \eta_E \varepsilon_{\rm SN} \dot{M}_*
    \; \; ,
\end{equation}
where $\varepsilon_{\rm SN}$ is the supernova energy output per unit mass of star formation and $\eta_E$ is the model's \textit{energy-loading parameter,} describing the fraction of supernova energy that couples with the CGM.

At the opposite extreme are uncoupled outflows that eject gas from a galaxy without transferring any of their energy or mass to the rest of the CGM. Whether or not the ejected gas returns to the galaxy depends on the outflow's specific energy
\begin{equation}
    \varepsilon_{\rm fb} = \frac {\eta_E \varepsilon_{\rm SN}} {\eta_M}
    \; \; .
\end{equation}
Paper I briefly explored such uncoupled outflows, showing that they do not return to the galaxy for $\varepsilon_{\rm fb} \gg \varepsilon_{\rm acc}$ and lead to excessive recycling for $\varepsilon_{\rm fb} \ll \varepsilon_{\rm acc}$. This paper will mostly focus on coupled outflows except where specifically indicated.

\subsection{Net Heating}

Radiative cooling enables accreted gas to accumulate in a halo's central galaxy.\footnote{Accreting baryons need to shed a specific energy similar to $\varepsilon_{\rm acc}$ before they can enter the central galaxy and remain there.} In Paper I, $\dot{M}_{\rm in}$ is the gas supply rate at which baryons pass from the CGM into the central galaxy's ISM, and $\dot{E}_{\rm loss}$ is the rate of atmospheric energy loss. With those losses included, the \textit{net} heating rate associated with star formation and supernova feedback is 
\begin{equation}
    \eta_E \varepsilon_{\rm SN} \dot{M}_* - \dot{E}_{\rm loss}
    \; \; .
\end{equation}
If $\dot{M}_{\rm in}$ is steady and interstellar gas forms stars on a relatively short timescale, then $\dot{M}_* \approx \dot{M}_{\rm in} / (1 + \eta_M)$, and the net heating rate can be expressed as
\begin{equation}
    \left[ \eta_E \varepsilon_{\rm SN} 
                - (1 + \eta_M) \varepsilon_{\rm loss} \right] \dot{M}_* 
    \; \; ,
\end{equation}
where $\varepsilon_{\rm loss} \equiv \dot{E}_{\rm loss} / \dot{M}_{\rm in}$.
Whether or not the supernova feedback loop results in a net gain or net loss of atmospheric energy therefore depends on how $\eta_E \varepsilon_{\rm SN}$ compares with $( 1 + \eta_M ) \varepsilon_{\rm loss}$.

\subsection{Galactic Gas Supply}

The minimalist regulator model assumes that $\dot{M}_{\rm in}$ depends on $E_{\rm CGM}$ and $M_{\rm CGM}$. Here we will also assume that $\dot{M}_{\rm in}$ depends on a potentially large vector of parameters $\mathbf{q} = (q_1 , q_2 , ... , q_N)$ describing the structure of the CGM, the cosmological accretion rate, and the halo's gravitational potential, so that the function representing the galaxy's gas supply can be expressed as
\begin{equation}
    \dot{M}_{\rm in} ( \, \varepsilon_{\rm CGM},f_{\rm CGM} \, | \, 
        \mathbf{q} \, )
        \; \; .
\end{equation}
When determining the feedback system's equilibrium states, Paper I assumes that
\begin{equation}
    \frac {\partial \dot{M}_{\rm in}} {\partial \varepsilon_{\rm CGM}} < 0
    \; \; ,
\end{equation}
so that increasing the atmosphere's specific energy reduces the galaxy's gas supply, as long as the model's parameters do not change.

\subsection{Equilibria}

Now the stage is set for describing the minimalist regulator model's equilibrium states. They are the states in which both $\varepsilon_{\rm CGM}$ and $f_{\rm CGM}$ remain constant, as long as the parameter set $\mathbf{q}$ remains unchanged. Under those conditions, the galaxy's star formation rate converges toward
\begin{equation}
    \dot{M}_* = \frac {\dot{M}_{\rm in}} {1 + \eta_M}
\end{equation}
as long as the star formation timescale $t_{\rm SF} = M_{\rm ISM} / \dot{M}_*$ of the ISM is sufficiently short. In such an equilibrium state, $f_{\rm CGM}$ stays constant as long as $\dot{M}_{\rm in}$ is proportional to $\dot{M}_{\rm halo}$. Paper I shows that the halo's stellar baryon fraction $f_* \equiv M_* / f_{\rm b} M_{\rm halo}$ then asymptotically approaches
\begin{equation}
    f_{*,{\rm asy}} = \frac {\varepsilon_{\rm eq} - \varepsilon_{\rm acc}}
        {\eta_E \varepsilon_{\rm SN} + \varepsilon_{\rm eq} 
            - (1 + \eta_M) \varepsilon_{\rm loss} }
            \; \; ,
\end{equation}
in which $\varepsilon_{\rm eq}$ is the equilibrium value of $\varepsilon_{\rm CGM}$.

There are two distinct families of equilibrium states. One has $\varepsilon_{\rm eq} > \varepsilon_{\rm acc}$ and corresponds to atmospheres that have expanded, because cumulative feedback energy input exceeds cumulative radiative losses. The other has $\varepsilon_{\rm eq} < \varepsilon_{\rm acc}$ and corresponds to atmospheres that have contracted, because cumulative radiative losses exceed cumulative feedback energy input.

\subsubsection{Expanded Atmospheres}
\label{sec:Expanded}

To see how an expanded atmosphere settles toward equilibrium, consider what happens if radiative losses are negligible. In that case supernova energy input can keep $\varepsilon_{\rm CGM}$ constant by adding energy at a rate that offsets the difference $\varepsilon_{\rm eq} - \varepsilon_{\rm acc}$ between the atmosphere's equilibrium specific energy and the specific energy of accreting baryons. The required energy input rate is 
\begin{equation}
    \eta_E \varepsilon_{\rm SN} \dot{M}_* 
      \, \approx \, (\varepsilon_{\rm eq} - \varepsilon_{\rm acc}) 
                    (f_{\rm b} \dot{M}_{\rm halo} - \dot{M}_*)
    \label{eq:ExpansionEquilibrium}
    \; \; .
\end{equation}
Significant reductions of the gas supply rate $\dot{M}_{\rm in}$ require the difference $\varepsilon_{\rm eq} - \varepsilon_{\rm acc}$ to be similar to $v_{\rm c}^2$. Paper I expresses the magnitude of that difference in terms of the quantity
\begin{equation}
    \xi ( \mathbf{q}) \equiv 
        \frac {\varepsilon_{\rm eq} - \varepsilon_{\rm acc}} 
              {v_{\rm c}^2}
              \; \; ,
\end{equation}
which is a dimensionless number of order unity that depends on the parameter set $\mathbf{q}$. When written in terms of $\xi$, the halo's asymptotic stellar baryon fraction becomes
\begin{equation}
    f_{*,{\rm asy}} \, = \, \frac {\xi v_{\rm c}^2} 
        {\eta_E \varepsilon_{\rm SN} + \xi v_{\rm c}^2 
            + [\varepsilon_{\rm acc} - (1 + \eta_M) \varepsilon_{\rm loss}]}
        \label{eq:fstar_asy_xi}
\end{equation}
in which the bracketed term can be neglected if radiative losses are small. Much of this paper is devoted to interpreting that result and assessing its implications.

\subsubsection{Contracted Atmospheres}
\label{sec:Contracted}

Interpreting the minimalist regulator model's predictions for contracted atmospheres is more challenging. Formally, the equilibrium state of a contracted atmosphere corresponds to a radiative loss rate that \textit{exceeds} feedback energy input, with energy input from cosmological accretion making up for the shortfall in feedback. However, such an equilibrium state is likely to be unstable with respect to variations in the parameter set $\mathbf{q}$ that alter both the radiative loss rate and the gas supply coming from a galaxy's atmosphere. 

Paper I did not explore such parameter variations. Instead, it focused on how atmospheric contraction boosts the asymptotic equilibrium value of $f_*$. Generally speaking, atmospheric contraction leads to greater star formation rates because it enhances the atmosphere's recycling rate. Contraction happens when the specific energy of feedback outflows $(\varepsilon_{\rm fb})$ is less than the mean specific energy of accreted gas ($\varepsilon_{\rm acc}$). The outflowing baryons tend to cycle back through the galaxy because they return to the atmosphere with less specific energy than they had when they originally entered the halo. Consequently, they augment the galaxy's gas supply and boost its equilibrium star formation rate.

According to Paper I, feedback outflows with $\varepsilon_{\rm fb}$ substantially less than $\varepsilon_{\rm acc}$ result in a crisis of excessive star formation, regardless of whether the outflows are coupled or uncoupled. Section \ref{sec:Simulations} discusses how the relationship between $\varepsilon_{\rm fb}$ and $\varepsilon_{\rm acc}$ informs interpretations of supernova feedback, as modeled in cosmological simulations of galaxy evolution.

\subsection{Fluctuations and Convergence}

At any given moment, feedback is pushing a galaxy and its atmosphere toward one of those equilibrium states. For example, consider a particular model for $\dot{M}_{\rm in}$ that depends on a parameter set $\mathbf{q}$. As long as $\mathbf{q}$ remains constant, star formation asymptotically converges toward a value of $\dot{M}_*$ that depends on $\mathbf{q}$.

However, the parameter vector $\mathbf{q}$ is likely to fluctuate in response to variations in cosmological accretion, CGM structure, the halo's gravitational potential, and perhaps other factors. The atmosphere's equilibrium state therefore also fluctuates, meaning that the atmosphere's trajectory in the $f_{\rm CGM}$--$\varepsilon_{\rm CGM}$ phase plane wanders around rather than steadily converging toward a single fixed point, as it does in Figure 5 of Paper I. Nevertheless, the trajectory remains close to the \textit{set} of fixed points corresponding to the \textit{range} of fluctuations in $\mathbf{q}$. Future papers in this series will explore those fluctuations and a galaxy's responses to them in more depth.

\subsection{Ejective or Preventative?}
\label{sec:Ejective?}

\begin{figure*}[t]
\begin{center}
  \includegraphics[width=0.99\textwidth]{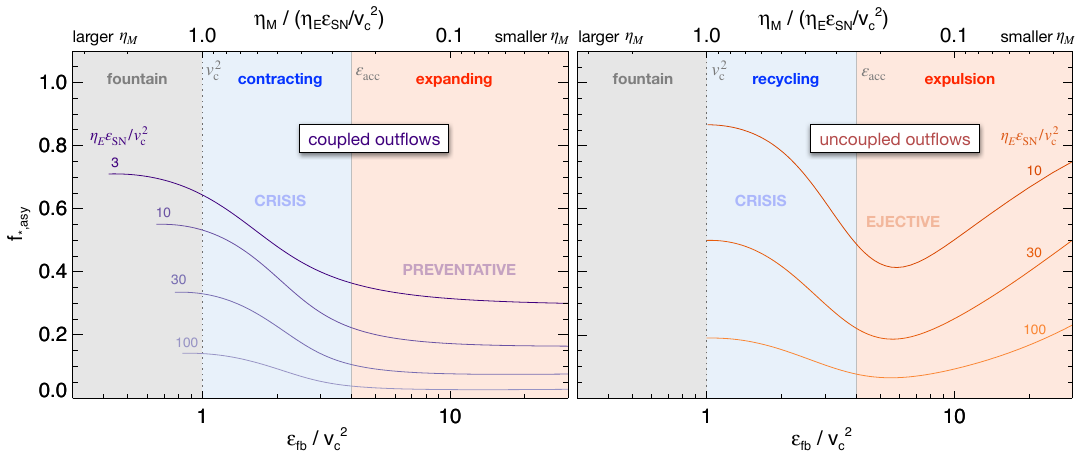}
\end{center}
\caption{Dependence of a halo's asymptotic stellar baryon fraction $f_{*,{\rm asy}}$ on the ratio $\varepsilon_{\rm fb}/ v_{\rm c}^2$ relating the specific energy of outflowing gas ($\varepsilon_{\rm fb} = \eta_E \varepsilon_{\rm SN} / \eta_M)$ to the halo's circular velocity. Left panel: Equilibrium states of Paper I's generic atmosphere model for coupled outflows, in which supernova feedback is primarily \textit{preventative.} Right panel: Equilibrium states of Paper I's ballistic circulation scenario for uncoupled outflows, in which supernova feedback is primarily \textit{ejective.} Numerical labels give values of $\eta_E \varepsilon_{\rm SN} / v_{\rm c}^2$ for each set of lines. Feedback models in the red region to the right of the solid vertical line marking $\varepsilon_{\rm fb} = \varepsilon_{\rm acc}$ expand a galaxy's atmosphere. Models in the blue region to the left of that vertical line allow a galaxy's atmosphere to contract. Models in the grey region to the left of the dotted line marking $\varepsilon_{\rm fb} = v_{\rm c}^2$ fail to propel gas very far out of a galaxy, resulting in a low-altitude fountain that quickly recycles ejected gas and does little to limit star formation.
}
\label{fig:fstar_epsej}
\end{figure*}

To conclude our discussion of Paper I's findings, we will consider their implications for how observations might be able to determine whether supernova feedback is primarily ejective or primarily preventative: 
\begin{enumerate}

    \item \textbf{Preventative.} Supernova feedback in Paper I's minimalist regulator model is primarily preventative, because the model assumes that supernova-driven feedback outflows couple with the CGM and add their energy to it. The left panel of Figure \ref{fig:fstar_epsej} shows how the model's star-formation predictions respond to mass loading, via the specific energy $\varepsilon_{\rm fb}$, using Paper I's ``generic atmosphere" formula for $\dot{M}_{\rm in}$. Purple lines represent how a halo's asymptotic stellar baryon fraction $f_{*,{\rm asy}}$ depends on the specific energy ratio $\varepsilon_{\rm fb}/v_{\rm c}^2$. Notice that they are nearly flat for $\varepsilon_{\rm fb} > \varepsilon_{\rm acc}$. In that regime, preventative supernova feedback reduces a galaxy's gas supply by expanding its atmosphere, making star formation insensitive to $\eta_M$ for reasons explained in Paper I and summarized in \S \ref{sec:Expanded}. However, the purple lines rapidly rise once $\varepsilon_{\rm fb}$ drops below $\varepsilon_{\rm acc}$, because then supernova feedback cannot prevent atmospheric contraction, which sharply increases the galaxy's gas supply and star formation rate.

    \item \textbf{Ejective.} The right-hand panel of Figure \ref{fig:fstar_epsej} presents a contrasting toy model from Paper I, the ``ballistic circulation" model, in which supernova-driven outflows are \textit{uncoupled}, meaning that they do not share their energy with the rest of the CGM. That feedback mode is purely ejective because it does nothing to prevent halo gas from entering the galaxy. Orange lines represent how the model's star-formation predictions depend on $\varepsilon_{\rm fb}$. All of those lines have a minimum near where $\varepsilon_{\rm fb} = \varepsilon_{\rm acc}$. Star formation increases toward greater values of $\varepsilon_{\rm fb}$ because galactic outflows eject less gas when $\eta_M$ is smaller. Star formation increases toward smaller values of $\varepsilon_{\rm fb}$ because of recycling: Ejected gas ultimately falls back into the galaxy, and reducing $\varepsilon_{\rm fb}$ shortens the recycling timescale, thereby increasing the galaxy's overall gas supply. Figure \ref{fig:fstar_epsej} therefore implies that ejective feedback needs to have $\varepsilon_{\rm fb} \approx \varepsilon_{\rm acc}$ in order to be optimally effective. 
    
\end{enumerate}
%Preventative feedback, on the other hand, effectively suppresses star formation for any value of $\varepsilon_{\rm fb}$ similar to or greater than $\varepsilon_{\rm acc}$, as long as the outflowing energy couples with the rest of the CGM. 

Comparing those two alternatives suggests that measurements showing how $\varepsilon_{\rm fb}$ depends on $v_{\rm c}^2$ might help distinguish between preventative and ejective feedback modes. Such measurements of both simulated and real galactic outflows need to include all forms of energy contributing to $\varepsilon_{\rm fb}$, including the thermal energy of a hot component that might be difficult to observe. Values of $\varepsilon_{\rm fb}$ consistently exceeding $\varepsilon_{\rm acc} \approx 4 v_{\rm c}^2$ would signify modes of feedback that need to be preventative in order to be effective. Observations consistently implying $\varepsilon_{\rm fb} \approx 4 v_{\rm c}^2$ would be more ambiguous, but a strong correlation between $\varepsilon_{\rm fb}$ and $v_{\rm c}^2$ could be interpreted as evidence favoring ejective feedback modes. 

%Figure \ref{fig:fstar_epsej} also suggests that the predictions of preventative feedback models with $\varepsilon_{\rm fb} > \varepsilon_{\rm acc}$ should be robust to variations in mass loading, because all values of $\eta_M$ significantly less than $\eta_E \varepsilon_{\rm SN} / \varepsilon_{\rm acc}$ result in similar amounts of star formation. Purely ejective feedback, on the other hand, has a sweet spot near $\eta_M = \eta_E \varepsilon_{\rm SN} / \varepsilon_{\rm acc}$ that minimizes star formation. Smaller proportions of mass loading concentrate feedback energy within a smaller proportion of the halo's baryons, ejecting them from the halo without significantly reducing the galaxy's gas supply. Larger proportions fail to prevent atmospheric contraction.

\section{Stellar Baryon Fractions}
\label{sec:Predictions}

This section discusses model predictions for the fraction $f_*$ of a halo's baryons contained within stars and compares them with observations of the $f_*$--$M_{\rm halo}$ relation. Section \ref{sec:SAMs} starts the discussion by outlining the landscape of semi-analytic models for galaxy evolution. Section \ref{sec:Carr} then focuses on the similarities and differences between the minimalist regulator model and one of its progenitors, the \citet{Carr_2023ApJ...949...21C} regulator model. Section \ref{sec:EnergyLoading} explains why a dependence of $\eta_E$ on $M_{\rm halo}$ is needed bring both the Carr et al.~model and the minimalist regulator model into alignment with observations. Section \ref{sec:fstar_peak} points out that the peak of the observed $f_*$--$M_{\rm halo}$ relation corresponds to where the minimalist regulator model would transition from expansion to contraction, if feedback energy were to come only from supernovae. Section \ref{sec:SigmaSFR} speculates about why the peak remains near $M_{\rm halo} \sim 10^{12} \, M_\odot$ over cosmic time and suggests that evolution of the surface density of star formation may compensate for changes in the relationship between $M_{\rm halo}$ and $v_{\rm c}$. 

\subsection{The Semi-Analytic Landscape} \label{sec:SAMs}

Traditional semi-analytic models for galaxy evolution track how a halo's baryons flow through different reservoirs \citep[early examples include][]{WhiteFrenk1991ApJ...379...52W,kauffmann93,cole96,somerville99}. They typically have the following features: 
\begin{enumerate}

    \item Accretion of baryons onto the CGM tracks $f_b \dot{M}_{\rm halo}$, where $\dot{M}_{\rm halo}$ is the total mass growth rate of the gravitationally bound halo. The gas is assumed to reach the halo's virial temperature when it accretes, and its initial composition is assumed to be primordial. 

    \item Radiative cooling enables CGM gas to accrete onto the ISM. Its cooling rate depends on the temperature and density profile of the CGM. Usually, the CGM is assumed to be isothermal, with a temperature $\propto v_{\rm c}^2$ and a density profile $\propto r^{-2}$ \citep[for discussions of other density profiles, see][]{lu11,monaco14,stevens17,hou19}.

    \item Star formation converts ISM gas into stars, which enrich the ISM with heavier elements (i.e. ``metals"). Some of those metals become locked into stars. 

    \item Galactic winds eject some of the ISM gas, along with some metals, into either the CGM or an ``ejected reservoir'' beyond the CGM. Enrichment of the CGM affects its cooling rate.

    \item Some of the ejected gas and metals may ``re-accrete'' from the ejected reservoir onto the CGM. 

\end{enumerate}

Traditional semi-analytical models usually do not comprehensively track energy flows among these reservoirs, but energy budget arguments sometimes do play a role. For example, in the L-galaxies model and many of its relatives \citep[e.g.,][]{guo11,henriques15,zhong23}, supernova energy is able to eject gas from the CGM reservoir. First, supernova energy added to some ISM gas must unbind it from the galaxy. The excess energy is then added to the CGM, where it can unbind some of the CGM gas, removing it from the CGM reservoir \citep[see also][]{lagos13,cousin15}.

Also, later generations of semi-analytic models include energy input from black-hole accretion \citep[e.g.,][]{croton06,somerville08}, which is assumed to produce jets that deposit energy in the CGM and can offset radiative cooling. However, any energy input exceeding the radiative cooling rate is simply lost, and does not modify the temperature or profile of the CGM, meaning that it does not impact the next timestep beyond a one-time suppression of cooling and accretion. 

This reliance on gas ejection to suppress star formation explains why most traditional semi-analytic models need to adopt extremely high mass loading factors to match observational constraints on stellar baryon fractions \citep[e.g.,][]{pandya20,Mitchell_2020MNRAS.494.3971M,Carr_2023ApJ...949...21C}. In order to reproduce observations of the $f_*$--$M_{\rm halo}$ relation, $\eta_M$ needs to depend strongly on $M_{\rm halo}$, becoming as large as $\eta_M \sim 10^2$ among low-mass halos. A few semi-analytic models have included preventative feedback algorithms that enable supernova energy to limit accretion onto the CGM and/or ISM \citep[e.g.,][]{lu15,hirschmann16,lu17,Pandya_2023ApJ...956..118P}. That approach naturally lowers the required mass-loading factors, but the ad hoc preventative feedback recipes are not self-consistently coupled to the atmosphere's energy budget.

In contrast, the progenitors of Paper I's minimalist regulator model explicitly track energy flows. \citet{Carr_2023ApJ...949...21C} track how energy input from both supernova feedback and cosmological accretion affects the CGM temperature and allow some of the CGM to escape if that temperature becomes too large. \citet{Pandya_2023ApJ...956..118P} include both of those energy sources and track evolution of both thermal and turbulent energy components in the CGM. However, neither \citet{Carr_2023ApJ...949...21C} nor \citet{Pandya_2023ApJ...956..118P} track gains or losses of gravitational potential energy as the CGM expands or contracts in response to changes in the balance between energy input and radiative cooling.

One semi-analytic model that \textit{does} track potential energy gains and losses was proposed by \citet{SharmaTheuns_2020MNRAS.492.2418S}. The details of their approach differ from the minimalist regulator model, but one of its outcomes is fundamentally similar. They find that star-formation rates at $M_{\rm halo} \lesssim 10^{12} \, M_\odot$ become linked to the halo's cosmological accretion rate, as in equation (\ref{eq:ExpansionEquilibrium}), because stellar feedback is compensating for cosmological accretion, not radiative cooling. 

\subsection{Predictions from Carr et al.}
\label{sec:Carr}

Now we will take a closer look at the semi-analytic regulator model of \citet{Carr_2023ApJ...949...21C}, who obtained $f_*$--$M_{\rm halo}$ predictions that were insensitive to the mass loading parameter $\eta_M$. They were also insensitive to heavy-element enrichment of the CGM. What mattered most was the energy loading parameter $\eta_E$.

\begin{figure}[t]
\begin{center}
  \includegraphics[width=0.50\textwidth]{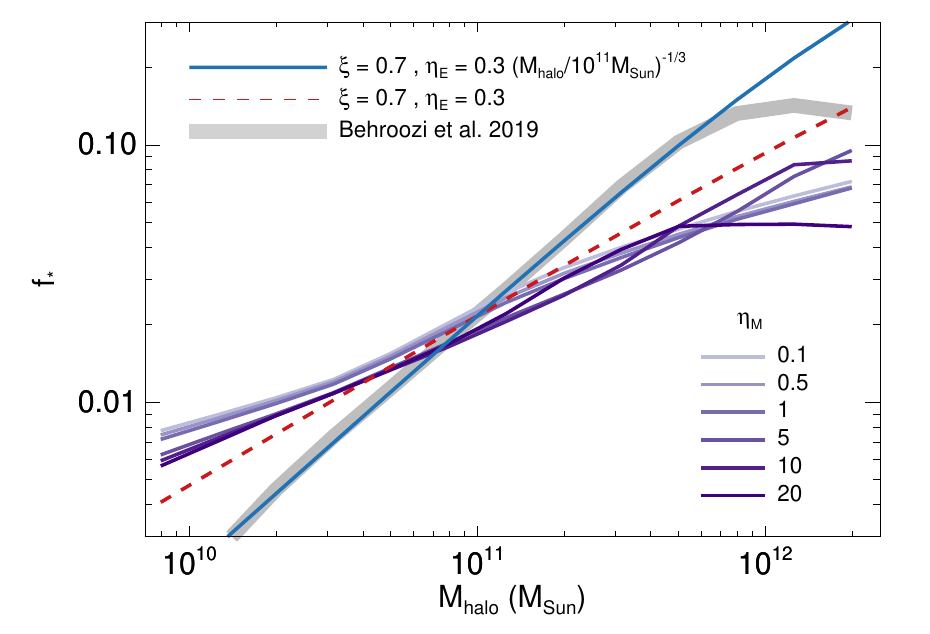}
\end{center}
\caption{Comparisons of predicted $f_*$--$M_{\rm halo}$ relations with observations. A grey band shows the low-redshift relation from \citet{Behroozi2019MNRAS.488.3143B}. Purple lines show predicted relations from \citet{Carr_2023ApJ...949...21C} for $\eta_E = 0.3$ and various values of $\eta_M$ given in the legend. A dashed red line shows the asymptotic stellar baryon fraction predicted by the minimalist regulator model for $\eta_E = 0.3$, $\xi = 0.7$, and negligible radiative losses. A solid blue line shows the minimalist regulator model's prediction for $\eta_E = 0.3 (M_{\rm halo}/10^{11} \, M_\odot)^{-1/3}$ and $\xi = 0.7$.
}
\label{fig:fstar_Carr}
\end{figure}

Figure~\ref{fig:fstar_Carr} illustrates how the $f_*$ predictions of  \citet{Carr_2023ApJ...949...21C} depend on $M_{\rm halo}$ and $\eta_M$. All of the purple lines represent models with $\eta_E = 0.3$ and values of $\eta_M$ ranging from 0.1 to 20. They result from integrating a regulator model similar to the minimalist regulator model of Paper I over cosmic time. Those lines are almost identical, showing that the predicted dependence of cumulative star formation on $\eta_M$ is generally weak, in alignment with the minimalist regulator model of Paper I, as expressed in equation (\ref{eq:fstar_asy_xi}) and as illustrated in the ``expanding" region of Figure \ref{fig:fstar_epsej}.

However, the Carr et al.~model differs from the minimalist regulator model in three important ways: 
\begin{enumerate}
    \item Baryons leave the CGM if atmospheric expansion pushes them beyond the halo's virial radius ($R_{\rm halo}$). 
    \item The model's treatment of gravitational potential energy is less complete.
    \item A galaxy's atmosphere cannot contract while radiative cooling exceeds feedback heating.
\end{enumerate}
The first difference opens another channel for atmospheric energy loss, because baryons going beyond $R_{\rm halo}$ carry away whatever specific energy they had before leaving, thereby reducing both $E_{\rm CGM}$ and $M_{\rm CGM}$. The second difference means that the specific energy of baryons entering the halo via cosmic accretion does not reflect all of their potential energy. Baryons accreting onto the CGM in the Carr et al.~model have a specific energy $\varepsilon_{\rm acc} = 0.66 v_{\rm c}^2$, and expansion of the CGM pushes baryons beyond $R_{\rm halo}$ and out of the CGM if $\varepsilon_{\rm CGM}$ exceeds $\varepsilon_{\rm acc}$. In contrast, accreting baryons in the minimalist regulator model are assumed to have $\varepsilon_{\rm acc} \approx 4 v_{\rm c}^2$, because $\varepsilon_{\rm acc}$ includes gravitational potential energy, in addition to thermal and kinetic energy. The third difference means that excess radiative cooling cannot trigger a runaway star formation crisis.

\subsection{Dependence of $\eta_E$ on $M_{\rm halo}$}
\label{sec:EnergyLoading}

A grey band in Figure \ref{fig:fstar_Carr} shows the $f_*$--$M_{\rm halo}$ relation that \citet{Behroozi2019MNRAS.488.3143B} inferred from observations. In order to fit that relation, \citet{Carr_2023ApJ...949...21C} had to assume that $\eta_E$ was a function of halo mass, becoming smaller as halo mass increases. 
Using a power-law relation with
\begin{equation}
    \eta_E = 0.26 \left( \frac {M_{\rm halo}} {10^{11} \, M_\odot} \right)^{-0.6}
\end{equation}
they were able to bring their predictions into good agreement with the observations. 

Contemporary cosmological simulations of galaxy evolution often do something similar, by employing parametric formulae that indirectly make the efficiency of supernova feedback a function of halo mass \citep[e.g.,][]{Schaye_EAGLE_2015MNRAS.446..521S,Nelson_outflows_2019MNRAS.490.3234N,Pillepich_TNG50_2018MNRAS.475..648P}. Each formula's parameters are tuned to reproduce the observable properties of galaxies. To obtain agreement, the simulations' equivalent of $\eta_E$ needs to become smaller as $M_{\rm halo}$ rises toward $\sim 10^{12} \, M_\odot$ \citep[e.g.,][]{CrainEAGLE+2015,Pillepich_2018MNRAS.473.4077P}.

A dashed red line in the figure shows the asymptotic $f_*$--$M_{\rm halo}$ relation predicted by the minimalist regulator model with $\xi = 0.7$, $\eta_M = 0.3$, and $\varepsilon_{\rm SN} = (700 \, {\rm km \, s^{-1}})^2$ when the radiative-loss correction in equation (\ref{eq:fstar_asy_xi}) is ignored. It is similar to the Carr et al.~predictions in the vicinity of $M_{\rm halo} \approx 10^{11} \, M_\odot$, but has a slightly steeper slope. At least two factors contribute to the difference in slope. First, early star formation in the low-mass halos of Carr et al.~happens under conditions far from equilibrium. Second, the equilibrium specific energy of the CGM among low-mass halos in Carr et al.~is significantly greater than $0.66 v_{\rm c}^2$. Consequently, the atmospheric energy losses sustained when expansion pushes gas beyond the virial radius are larger than in the minimalist regulator model. More star formation (or larger $\eta_E$) is then required to drive a given amount of atmospheric expansion. 

A simple power-law dependence of $\eta_E$ on $M_{\rm halo}$, shallower than the one used by Carr et al.,~brings the asymptotic $f_*$--$M_{\rm halo}$ relation from the minimalist regulator model into good agreement with the observations. The blue line in Figure \ref{fig:fstar_Carr} shows the result of using
\begin{equation}
    \eta_E = 0.3 \left( \frac {M_{\rm halo}} {10^{11} \, M_\odot} \right)^{-1/3}
        \label{eq:etaE_fit}
\end{equation}
instead of $\eta_E = 0.3$. It matches the observations well, except near the peak of the observed $f_*$--$M_{\rm halo}$ relation, which we will discuss next. Note that any set of parameter choices with $\eta_E \propto M_{\rm halo}^{-1/3}$ and $\xi / \eta_E \approx 0.7 / 0.3$ near $M_{\rm halo} = 10^{11} \, M_\odot$ would give similar results, because the parameters $\xi$ and $\eta_M$ are nearly degenerate.

\subsection{Atmospheric Contraction and the $f_*$ Peak}
\label{sec:fstar_peak}

Next we consider the well-known peak in the $f_*$--$M_{\rm halo}$ relation, observed to persist near $M_{\rm halo} \sim 10^{12} M_\odot$ across most of cosmic time \citep[e.g.,][]{Behroozi2019MNRAS.488.3143B}. None of the models shown in Figure \ref{fig:fstar_Carr} reproduces it, which is not surprising because those models do not include feedback resulting from accretion onto a central supermassive black hole. Nevertheless, the mimimalist regulator model provides a clue about \textit{why} black-hole feedback becomes important as $M_{\rm halo}$ approaches $10^{12} \, M_\odot$. Apparently, that is where supernova feedback starts to fail, permitting an atmospheric contraction crisis to fuel strong black-hole feedback \citep[see also][]{SharmaTheuns_2020MNRAS.492.2418S}.

\subsubsection{The Expansion/Contraction Border}

According to the minimalist regulator model, there is a borderline value of halo circular velocity at which the specific energy of supernova-driven outflows ($\varepsilon_{\rm fb}$) is close to the specific energy of gas accreting onto the halo ($\varepsilon_{\rm acc}$). Below the $v_{\rm c}$ borderline, supernova feedback causes expansion of a galaxy's atmosphere. Above the $v_{\rm c}$ borderline, supernova feedback cannot prevent atmospheric contraction, because outflows with $\varepsilon_{\rm fb} \ll \varepsilon_{\rm acc}$ tend to \textit{reduce} the mean specific energy of the CGM. 

Paper I assumed $\varepsilon_{\rm acc} \approx 4 v_{\rm c}^2$, because that is the typical difference between a halo's gravitational potential at the virial radius and its potential near the central galaxy's effective radius. The borderline should therefore be near
\begin{equation}
    v_{\rm c} \approx \frac {\varepsilon_{\rm fb}^{1/2}} {2}
              \approx 350 \, {\rm km \, s^{-1}}
        \times \left( \frac {\eta_E} {\eta_M} \right)^{1/2}
        \label{eq:Borderline1}
\end{equation}
for $\varepsilon_{\rm SN} \approx (700 \, {\rm km \, s^{-1}})^2$. Plausible estimates of $\eta_E$ and $\eta_M$ place the Milky Way near that borderline. For example, extrapolating equation (\ref{eq:etaE_fit}) to $M_{\rm halo} = 10^{12} \, M_\odot$ gives $\eta_E = 0.14$, putting the borderline at
$v_{\rm c} \approx 140 \, \eta_M^{-1/2} \, {\rm km \, s^{-1}}$, which reduces to $v_{\rm c} \approx 200 \, {\rm km \, s^{-1}}$ for $\eta_M \approx 0.5$.

The counterpart to that borderline in the \citet{Carr_2023ApJ...949...21C} models manifests in Figure~\ref {fig:fstar_Carr} as a break in the $f_*$--$M_{\rm halo}$ relation.  You can see that the lines for $\eta_M = 10$ and 20 both break near $10^{12} \, M_\odot$, becoming nearly flat toward larger $M_{\rm halo}$. The locations of those breaks disagree with equation (\ref{eq:Borderline1}) because the Carr et al.~models assume $\varepsilon_{\rm acc} \approx 0.66 v_{\rm c}^2$ instead of $\varepsilon_{\rm acc} \approx 4 v_{\rm c}^2$. That is why the line for $\eta_M = 10$ breaks near $v_{\rm c} \approx 156 \, {\rm km \, s^{-1}}$ and the line for $\eta_M = 20$ breaks near $v_{\rm c} \approx 110 \, {\rm km \, s^{-1}}$. 

Beyond those break points, supernova feedback in the Carr et al.~models falls short of replacing the initial specific energy of baryons that have cycled through the galaxy, and so it cannot push halo gas beyond the virial radius. However, like many other semi-analytic models for galaxy evolution, the Carr et al.~model does not allow a galaxy's atmosphere to contract. Consequently, runaway cooling cannot increase the galaxy's gas supply. Instead, the gas supply saturates at a particular value of $\dot{M}_{\rm in} / \dot{M}_{\rm acc}$, and star formation saturates at a particular value of $f_*$ that is proportional to $1/(1 + \eta_M)$, declining as $\eta_M$ increases. 

\subsubsection{Contraction and Black Hole Feedback}

The main thing we are trying to illustrate through this particular comparison with \citet{Carr_2023ApJ...949...21C} is the importance of including both gravitational potential energy and atmospheric contraction in future semi-analytic models for galaxy evolution. Gravitational potential energy is important for setting the borderline at which $\varepsilon_{\rm fb} \approx \varepsilon_{\rm acc}$. Allowing for atmospheric contraction is \textit{essential} for modeling feedback that has $\varepsilon_{\rm fb} < \varepsilon_{\rm acc}$.

Cosmological simulations seem consistent with the idea that strong black-hole feedback is a galaxy's response to the atmospheric contraction that the minimalist regulator model predicts for $M_{\rm halo} \gtrsim 10^{12} \, M_\odot$. For example, the analyses of \citet{Ayromlou_2023MNRAS.524.5391A} and \citet{Wright2024}, which compare IllustrisTNG, EAGLE, and SIMBA, show that atmospheric expansion generally lessens as halo mass increases. However, the trend reverses near $M_{\rm halo} \approx 10^{12} \, M_\odot$ in IllustrisTNG and SIMBA, and becomes flatter there in EAGLE, because that is where black-hole feedback becomes the dominant mode in all three simulations. Furthermore, the ``closure radius" for $10^{12} \, M_\odot$ halos at $z \approx 2$ is close to $R_{\rm halo}$ in both IllustrisTNG and SIMBA, indicating that the atmospheres of those halos are on the borderline between expansion and contraction.

\subsection{$\Sigma_{\rm SFR}$ and the Borderline}
\label{sec:SigmaSFR}

So far, we have been discussing the peak in the $f_*$--$M_{\rm halo}$ relation as if it were a borderline in $v_{\rm c}$ instead of $M_{\rm halo}$. However, higher-redshift galaxies at fixed $M_{\rm halo}$ have greater $v_{\rm c}$. To remain near $M_{\rm halo} \sim 10^{12} M_\odot$, the borderline for atmospheric contraction therefore needs to be at larger $v_{\rm c}$ earlier in time. This section suggests that evolution of the surface density $\Sigma_{\rm SFR}$ of a galaxy's star formation rate may be what keeps the borderline for atmospheric contraction close to $M_{\rm halo} \sim 10^{12} M_\odot$. 

Small-box simulations designed to resolve the ISM physics at the root of supernova feedback \citep[e.g.,][]{Kim_2017ApJ...834...25K,LiBryanOstriker_2017ApJ...841..101L,Fielding_2018MNRAS.481.3325F,Armilotta_2019MNRAS.490.4401A} have explored how $\Sigma_{\rm SFR}$ is related to $\eta_E / \eta_M$. In the compilation of such simulations presented by \citet{LiBryan2020}, the best fit for hot outflows yields 
\begin{equation}
    \frac {\eta_E} {\eta_M}
        = 0.78 \left( \frac {\Sigma_{\rm SFR}}
                            {M_\odot \, {\rm kpc^{-2} 
                            \, yr^{-1}}}
               \right)^{0.2}
    \label{eq:HotOutflows}
    \; \; .
\end{equation}
Inserting that fit into equation (\ref{eq:Borderline1}) gives the borderline criterion
\begin{equation}
    v_{\rm c} 
        \approx 310 \, {\rm km \, s^{-1}}
                \left( \frac {\Sigma_{\rm SFR}}
                            {M_\odot \, {\rm kpc^{-2} 
                            \, yr^{-1}}}
               \right)^{0.1}
        \label{eq:Borderline2}
            \; \; ,
\end{equation}
which implies that the tipping point for atmospheric contraction may depend weakly on the surface density of star formation within a halo's central galaxy. For the Milky Way, observations within the solar circle show that $\Sigma_{\rm SFR} \approx 10^{-2} \, M_\odot \, {\rm kpc^{-2} \, s^{-1}}$ \citep{Elia_2022_SigmaSFR}. Equation (\ref{eq:HotOutflows}) then gives $\eta_E / \eta_M \approx 0.3$, and equation (\ref{eq:Borderline2}) puts the borderline between expansion and contraction near $200 \, {\rm km \, s^{-1}}$, in accord with the Milky Way estimate in \S \ref{sec:fstar_peak}.

Higher-redshift galaxies with stellar masses and halo masses similar to the Milky Way's have much greater star formation rates but somewhat smaller effective radii, and therefore have much greater $\Sigma_{\rm SFR}$ values \citep[e.g.][]{Salim_SigmaSFR_2023ApJ...958..183S}. More highly concentrated star formation could therefore be what keeps the peak of the $f_*$--$M_{\rm halo}$ relation near $10^{12} \, M_\odot$. For example, consider a high-redshift galaxy with $\Sigma_{\rm SFR} \sim 1 \, M_\odot \, {\rm kpc^{-2} \, yr^{-1}}$. According to equation (\ref{eq:Borderline2}), its borderline circular velocity is near $300 \, {\rm km \, s^{-1}}$, corresponding to $M_{\rm halo} \sim 3 \times 10^{12} \, M_\odot$ at $z \sim 2$. In other words, the borderline above which supernova feedback cannot prevent atmospheric contraction around high-redshift galaxies is at greater $v_{\rm c}$, but the halo mass at a given value of $v_{\rm c}$ is smaller. Further investigation will be needed to determine whether those two countervailing effects do indeed combine to keep the peak of $f_*$ near $10^{12} \, M_\odot$.

\section{Interpretation of Simulations}
\label{sec:Simulations}

We anticipate that the minimalist regulator model and its phase-plane trajectories will be useful tools for interpreting cosmological simulations of galaxy evolution and comparing them with observational data. This section suggests two potential applications to be pursued in future work: (1) using phase-plane trajectories to analyze how different feedback models shape the global characteristics of galactic atmospheres, and (2) analyzing how the outcomes of cosmological simulations depend on the specific energy of the galactic outflows they produce. 

\subsection{Phase-Plane Trajectories}

Any given simulation contains a population of halos with measurable values of $M_{\rm CGM}$, $E_{\rm CGM}$, and $M_{\rm halo}$. Those measurements can be converted to $f_{\rm CGM} = M_{\rm CGM} / f_{\rm b} M_{\rm halo}$ and $\varepsilon_{\rm CGM} = E_{\rm CGM} / M_{\rm CGM}$ and plotted in the $f_{\rm CGM}$--$\varepsilon_{\rm CGM}$ phase plane. Mapping the flow of those points through the phase plane would then reveal how the atmospheres of galaxies respond to feedback. Also, the two-dimensional distribution of the phase-plane points at a particular moment can be compared with both observational constraints and the predictions of semi-analytic models \citep[see][for examples]{pandya20,Pandya_2023ApJ...956..118P}. 

Applying the comprehensive accounting approach adopted in Paper I to particle-based simulations should start with identification of baryonic particles that were originally cospatial with a halo's current dark matter particles. Summing the masses of baryonic particles that did not form stars and are not currently part of some galaxy's interstellar medium then gives $M_{\rm CGM}$. However, assessing $E_{\rm CGM}$ requires another step beyond a simple sum of thermal and non-thermal particle energy, because the dominant contribution to $E_{\rm CGM}$ usually comes from gravitational potential energy. Comprehensive accounting algorithms to determine $E_{\rm CGM}$ therefore need to identify a halo's center, calculate the gravitational potential with respect to that centroid, and designate an appropriate zero point.

Methods for measuring $M_{\rm CGM}$ and $E_{\rm CGM}$ in grid-based codes cannot be as direct, because baryons in those simulations do not retain a traceable identity. An alternative method for estimating $M_{\rm CGM}$ would be to measure the sum of $M_*$ and $M_{\rm ISM}$ within a halo and subtract it from $f_{\rm b} M_{\rm halo}$. The atmosphere's total energy can then be estimated from the intergalactic baryons within a sphere containing gas mass $M_{\rm CGM}$. Summing the thermal, non-thermal, and gravitational potential energies of those atmospheric baryons yields an estimate of $E_{\rm CGM}$. However, that sum may underestimate the total amount of feedback energy added to the atmosphere if a significant fraction of the halo's original baryons have been ejected far beyond that sphere and replaced by baryons falling toward the halo for the first time.

Obtaining observational constraints on $M_{\rm CGM}$ and $E_{\rm CGM}$ may be more challenging, because baryons near or beyond a halo's virial radius are so difficult to detect. However, the distribution of atmospheric mass as a function of radius places a lower limit on $E_{\rm CGM}$ that can be useful for distinguishing among various feedback models. One possible proxy for $E_{\rm CGM}$ is the circumgalactic baryon fraction within the virial radius 
\begin{equation}
    f_{\rm CGM,vir} = \frac {M_{\rm CGM}(< R_{\rm halo})} {f_{\rm b} M_{\rm halo}}
    \; \; .
\end{equation}
\citep[for comparisons of $f_{\rm CGM,vir}$ among simulations, see][]{Wright2024}. Another is the closure radius \citep[e.g.,][]{Ayromlou_2023MNRAS.524.5391A}.

\subsection{Specific Energy and its Consequences}

Perhaps more fundamentally, the minimalist regulator model clarifies how the \textit{specific energy} of a feedback outflow determines the ultimate outcome of a cosmological galaxy-evolution simulation. According to Figure \ref{fig:fstar_epsej}, supernova-driven outflows that couple with the CGM result in stellar baryon fractions that are insensitive to mass loading, as long as $\varepsilon_{\rm fb}$ exceeds $\varepsilon_{\rm acc}$. In that regime, a central galaxy's long-term star formation rate becomes closely linked to the mass-accretion rate of its halo, because supernova feedback is mainly balancing baryon accretion, not radiative cooling. A halo's eventual stellar baryon fraction therefore depends mostly on the \textit{fraction} of supernova energy that couples with the CGM ($\eta_E$), because $\eta_E \varepsilon_{\rm SN} M_*$ is always comparable to $f_{\rm b} M_{\rm halo} v_{\rm c}^2$ \citep[see also][]{SharmaTheuns_2020MNRAS.492.2418S}.

When supernova feedback is balancing cosmological accretion, raising $\eta_M$ reduces atmospheric expansion, because more of the supernova energy going into the CGM is lost to radiative cooling (see Paper I for a more detailed explanation). The effect is minor until $\eta_M$ becomes large enough for $\varepsilon_{\rm fb}$ to fall below $\varepsilon_{\rm acc}$. When those two specific energies are comparable, then a galaxy's atmosphere does not expand, because radiative cooling sheds all the energy input from supernovae and cosmological accretion. 
%But when $\varepsilon_{\rm fb}$ from supernova feedback alone are smaller than $\varepsilon_{\rm acc}$, the galaxy's atmosphere inevitably contracts, perhaps triggering strong outbursts of black-hole feedback having much greater values of $\varepsilon_{\rm fb}$.

To point the way toward more detailed comparisons with simulations, this concluding section takes a tiny step in that direction, briefly sketching out where the Illustris, EAGLE, and IllustrisTNG simulations reside within the parameter space of the minimalist regulator model. What it presents helps to explain how those simulations manage to produce galaxies with similar star-formation properties but strikingly different atmospheric characteristics.

\begin{figure*}[t]
\begin{center}
  \includegraphics[width=0.7\textwidth]{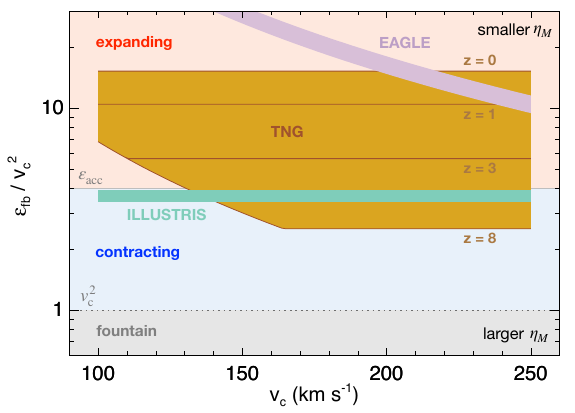}
\end{center}
\caption{Approximate ranges of specific supernova feedback energy ($\varepsilon_{\rm fb} = \eta_E \varepsilon_{\rm SN} / \eta_M$) implemented in cosmological numerical simulations. Each colored band shows
how the scaled specific energy $\varepsilon_{\rm fb} / v_{\rm c}^2$ depends on the approximate circular velocity $v_{\rm c}$ of a galaxy's halo in the range 100--250 ${\rm km \, s^{-1}}$. A horizontal blue-green stripe represents the original Illustris simulations. A diagonal purple stripe represents EAGLE. The gold region represents the IllustrisTNG simulations, in which $\varepsilon_{\rm fb} / v_{\rm c}^2 \propto H^{-2/3}(z)$, and labeled brown stripes show $\varepsilon_{\rm fb} / v_{\rm c}^2$ at particular values of redshift $z$. The red, blue, and grey shaded regions correspond to the expanding, contracting, and fountain regions of Figure \ref{fig:fstar_epsej}. According to Paper I's minimalist regulator model, coupled supernova-driven outflows with greater values of $\varepsilon_{\rm fb}/v_{\rm c}^2$ should result in more expansion of the CGM without changing a halo's stellar baryon very much, as long as $\varepsilon_{\rm fb}$ significantly exceeds $\varepsilon_{\rm acc}$. The CGM properties of low-mass halos in Illustris, EAGLE, and IllustrisTNG qualitatively align with that prediction. 
}
\label{fig:vc_epsfb_plot}
\end{figure*}

\subsubsection{Illustris: $\varepsilon_{\rm fb} \approx \varepsilon_{\rm acc}$}

We will start with Illustris, because it illustrates what happens when $\varepsilon_{\rm fb}$ is comparable to $\varepsilon_{\rm acc}$. In Illustris, a galactic wind's radial outflow speed ($v_{\rm w}$) is directly linked to the one-dimensional velocity dispersion of dark-matter particles in the galaxy's neighborhood ($\sigma_{\rm 1D}$) through the equation
\begin{equation}
    v_{\rm w} = \kappa_{\rm w} \sigma_{\rm 1D}
    \; \; . 
\end{equation}
The fiducial value of the proportionality constant is $\kappa_{\rm w} = 3.7$ \citep{Vogelsberger_IllustrisModel_2013MNRAS.436.3031V}. This constant determines the mass-loading factor of Illustris winds, which is proportional to the specific supernova energy output divided by $v_{\rm w}^2$.

Galactic winds passing from the ISM into the CGM in Illustris therefore start with a specific energy at least as great as
\begin{equation}
    \varepsilon_{\rm fb} 
        \: \approx \: \frac {v_{\rm w}^2} {2}
        \: \approx \: 3.4 v_{\rm c}^2
    \; \; ,
\end{equation}
assuming that $v_{\rm c}^2 \approx 2 \sigma_{\rm 1D}^2$. The total specific energy of ejected gas particles may be a little greater if the gas starts with a rotational velocity component similar to $v_{\rm c}$. Figure \ref{fig:vc_epsfb_plot} therefore represents Illustris with a blue-green stripe with a width spanning the interval $3.4 \leq \varepsilon_{\rm fb} / v_{\rm c}^2 \leq 3.9$. 

Notably, the specific energy of supernova-driven Illustris outflows is close to the specific energy of accreting baryons ($\varepsilon_{\rm acc} \approx 4 v_{\rm c}^2$), implying that supernova feedback alone cannot significantly expand a galaxy's atmosphere in the fiducial Illustris simulations. That implication is consistent with the unusual baryon richness of low-mass halos in Illustris, which have $f_{\rm CGM,vir} \gtrsim 1$ at $M_{\rm halo} \sim 10^{11} \, M_\odot$ \citep[e.g.,][]{Pillepich_2018MNRAS.473.4077P}. In contrast, the higher-mass Illustris halos ($M_{\rm halo} \sim 10^{11.5}$--$10^{13.5} \, M_\odot$) have values of $f_{\rm CGM,vir}$ up to an order of magnitude smaller because black-hole feedback is much stronger than supernova feedback in that mass range, and it greatly expands the atmospheres of the higher-mass halos.

Furthermore, numerical experiments with Illustris demonstrate that reducing $\eta_M$ below its fiducial value \textit{suppresses} long-term star formation. That outcome qualitatively aligns with the general trend traced by the purple lines in Figure \ref{fig:fstar_epsej}. For example, the ``faster winds" simulation presented in \citet{Vogelsberger_IllustrisModel_2013MNRAS.436.3031V} increases the $\kappa_{\rm w}$ parameter to 7.4 without changing the total amount of supernova energy injection, thereby reducing $\eta_M$ by a factor of 4. In the parameter space of Figures \ref{fig:fstar_epsej} and \ref{fig:vc_epsfb_plot}, that change raises $\varepsilon_{\rm fb}$ to $\approx 4 \varepsilon_{\rm acc}$, and it results in substantially less star formation across all halo masses \citep[see the right panel of Figure 7 in][]{Vogelsberger_IllustrisModel_2013MNRAS.436.3031V}.

This qualitative trend appears to be inconsistent with the orange lines representing uncoupled outflows in the right panel of Figure \ref{fig:fstar_epsej}. Apparently, the ``faster winds" in Illustris are suppressing star formation more effectively than the fiducial winds, even though they eject less galactic gas, because they strongly couple with the CGM. To be clear, the Illustris winds, by construction, do not directly couple with a galaxy's ISM, and so the ``faster winds" must suppress star formation more effectively than the fiducial winds because they reduce the rate at which CGM gas flows into galaxies.

\subsubsection{EAGLE: $\varepsilon_{\rm fb} \gg \varepsilon_{\rm acc}$}

The next exhibit is EAGLE, in which the specific energy of a supernova-driven galactic outflow is larger than in Illustris and results in more extended atmospheres, with comparatively low values of $f_{\rm CGM,vir}$ \citep[e.g.,][]{Mitchell_2020MNRAS.494.3971M,Wijers_2020MNRAS.498..574W,Wright2024}.
For example, low-mass EAGLE halos have $f_{\rm CGM,vir} \sim 0.1$ at $\sim 10^{11} \, M_\odot$ and $f_{\rm CGM,vir} \sim 0.2$ at $\sim 10^{12} \, M_\odot$, indicating that feedback has substantially expanded their atmospheres.

To represent EAGLE in the parameter space of Figure \ref{fig:vc_epsfb_plot}, we use a purple stripe determined by the equation
\begin{equation}
    \varepsilon_{\rm fb}
        \: = \: \frac {3 k \Delta T} {2 \mu m_p} 
        \label{eq:EAGLE_fb}
        \; \; ,
\end{equation}
where $\mu m_p$ is the mean mass per gas particle. In EAGLE, the temperature increment $\Delta T = 10^{7.5} \, {\rm K}$ is great enough for supernova-driven winds to develop before radiative cooling drains too much of their thermal energy. As supernovae explode, the temperature of the ISM gas particles they heat rises by $\Delta T$, and the total amount of supernova energy input determines the \textit{number} of particles receiving temperature boosts. Consequently, $\Delta T$ determines the \textit{specific energy} of EAGLE's galactic winds and may be uncorrelated with the total feedback power flowing from the ISM into the CGM.

We have chosen to assume in equation (\ref{eq:EAGLE_fb}) and Figure \ref{fig:vc_epsfb_plot} that the gas particles in EAGLE's supernova-driven winds have specific energies that remain unchanged from the time they are heated to the time they exit the galaxy. It is possible for both radiative cooling and entrainment of additional mass to reduce the specific energy of an EAGLE outflow below the value in equation (\ref{eq:EAGLE_fb}) before it enters the CGM. However, the purple stripe representing EAGLE remains well above the expansion/contraction transition at $\varepsilon_{\rm fb} \approx \varepsilon_{\rm acc}$ among halos with circular velocities up to $250 \, {\rm km \, s^{-1}}$, meaning that the effectiveness of supernova feedback in such halos should be insensitive to the exact value of $\varepsilon_{\rm fb}$, as long as those outflows couple with the CGM. %And indeed, \citep{MitchellSchaye_2022MNRAS.511.2948M} show that the rate at which EAGLE feedback pushes gas beyond a halo's virial radius greatly exceeds the rate at which it pushes gas out of the halo's central galaxy.

According to Paper I and \S \ref{sec:Equilibrium}, stellar baryon fractions within such halos should depend primarily on $\eta_E \varepsilon_{\rm SN}$ and hardly at all on $\eta_M$. EAGLE achieves good agreement with observed $f_*$--$M_{\rm halo}$ relations by tuning a parameter that determines the proportion of supernova energy added to ISM gas \citep{CrainEAGLE+2015}. That form of tuning is conceptually similar to tuning the value of $\eta_E$ in the minimalist regulator model. However, the counterpart to $\eta_E$ in EAGLE needs to be greater than unity among low-mass halos to obtain satisfactory agreement, presumably to account for excessive unresolved cooling, and possibly also additional forms of feedback energy coming from the stellar population.

\citet{Mitchell_2020MNRAS.494.3971M} and \citet{MitchellSchaye_2022MNRAS.511.2948M} have analyzed the outflows from EAGLE galaxies and have explored how they shape the $f_*$--$M_{\rm halo}$ relation. The central galaxies of $10^{12} \, M_\odot$ halos have outflows with $\eta_M \sim 1$--2. In lower-mass halos, the central outflows have $\eta_M \propto v_{\rm c}^{-3/2}$. In all EAGLE halos, the mass outflow rates through $R_{\rm halo}$ exceed outflow rates from the central galaxy by a factor of several across most of cosmic time, demonstrating that galactic outflows in EAGLE do indeed cause atmospheric expansion by coupling with circumgalactic gas \citep[see also][]{Wright2024}. Interestingly, \citet{Mitchell_2020MNRAS.494.3971M} also compare the mass-loading factors used in several different semi-analytic models with EAGLE and show that those models employ values of $\eta_M$ that can differ by orders of magnitude, from both EAGLE and each other. Yet those models all produce similar predictions for the $f_*$--$M_{\rm halo}$ relation.

\subsubsection{IllustrisTNG: Evolving $\varepsilon_{\rm fb}/\varepsilon_{\rm acc}$}

To finish, we will consider IllustrisTNG, in which $\varepsilon_{\rm fb}$ at fixed $v_{\rm c}^2$ evolves with time and has a lower limit. More specifically, the speeds of galactic winds in IllustrisTNG are determined by
\begin{equation}
    v_{\rm w} = \max \left[ \kappa_{\rm w} \sigma_{\rm 1D}
                    \left( \frac {H_0} {H(z)} \right)^{1/3} 
                    , \: v_{\rm w,min} \right]
\end{equation}
in which the fiducial parameter values are $\kappa_{\rm w} = 7.4$ and $v_{\rm w,min} = 350 \, {\rm km\, s^{-1}}$ \citep{Pillepich_2018MNRAS.473.4077P}. The ratio $H(z)/H_0$ specifies how the Hubble expansion parameter $H$ at redshift $z$ compares with its current value ($H_0$) and keeps $v_{\rm w}$ at fixed \textit{halo mass} nearly constant. 

The gold region in Figure \ref{fig:vc_epsfb_plot} represents IllustrisTNG, and the brown stripes show how 
\begin{equation}
    \varepsilon_{\rm fb} = \frac {v_{\rm w}^2} {2 (1 - \tau_{\rm w})} 
\end{equation}
changes with redshift, for the fiducial thermal energy fraction $\tau_{\rm w} = 0.1$. You can see that the mass loading factors of supernova-driven outflows in IllustrisTNG halos with $v_{\rm c} \lesssim 200 \, {\rm km \, s^{-1}}$ lie in between their Illustris and EAGLE counterparts throughout most of cosmic time. At redshift $z \gtrsim 5$, the specific energy of a supernova-driven outflow in an IllustrisTNG halo with $v_{\rm c} \gtrsim 130 \, {\rm km \, s^{-1}}$ is similar to or less than in an Illustris halo with similar $v_{\rm c}$ and is also less than $\varepsilon_{\rm acc}$. However, the specific energies of the IllustrisTNG outflows increase as time proceeds, eventually becoming great enough for $\varepsilon_{\rm fb}$ to exceed $\varepsilon_{\rm acc}$. After that happens, supernova feedback in IllustrisTNG is able to expand the atmospheres of those halos. Also, IllustrisTNG feedback is \textit{always} capable of expanding galactic atmospheres in halos with $v_{\rm c} \lesssim 130 \, {\rm km \, s^{-1}}$, because of the lower limit it imposes on the outflow speed.

The baryon mass fractions within low-mass IllustrisTNG halos are less than the cosmic mean value ($f_{\rm b}$) at $z \approx 0$, even though black-hole feedback is negligible in that population \citep{Pillepich_2018MNRAS.473.4077P}. Also, the baryon fractions within EAGLE halos are typically 2 to 3 times smaller, indicating that a greater proportion of the baryons has been pushed beyond the virial radius \citep{MitchellSchaye_2022MNRAS.511.2948M,Wright2024}. In contrast, Illustris halos at $\sim 10^{11} \, M_\odot$ are baryon complete, as mentioned earlier. 

Those findings are qualitatively consistent with the relative locations of the Illustris, EAGLE, and IllustrisTNG bands in Figure \ref{fig:vc_epsfb_plot}. Central galaxies in low-mass IllustrisTNG halos have outflows with great enough specific energy to push out some of the halo gas, but they have less specific energy than EAGLE outflows in halos of similar mass. The EAGLE outflows are therefore capable of pushing away a greater proportion of the halo's baryons, whereas supernova feedback in Illustris cannot push away any of the halo's baryons.

Variants of the IllustrisTNG feedback algorithm presented in \citet{Pillepich_2018MNRAS.473.4077P} provide additional support for this interpretation of specific energy's role. For example, removing the $350 \, {\rm km \, s^{-1}}$ floor on wind speed results in several times more star formation within $10^{11} \, M_\odot$ halos and significantly more circumgalactic gas, presumably because $\varepsilon_{\rm fb}$ is smaller than $\varepsilon_{\rm acc}$ early in the histories of those halos. Also, the ``warmer winds" variant of IllustrisTNG boosts $\varepsilon_{\rm fb}$ by a factor of 2 without changing the supernova power output and results in more atmospheric expansion (i.e. smaller $f_{\rm CGM,vir}$) among low-mass halos ($< 10^{12} \, M_\odot)$.

\subsection{The Bottom Line}

Our overall point here is that the \textit{specific energy} of supernova-driven outflows ($\varepsilon_{\rm fb} = \eta_E \varepsilon_{\rm SN} / \eta_M$) and the \textit{total energy} of those outflows ($\eta_E \varepsilon_{\rm SN} M_*$) play different roles. According to the minimalist regulator model, outflows capable of expanding a galaxy's atmosphere self-regulate so that $\eta_E \varepsilon_{\rm SN} M_*$ approaches $\approx f_{\rm b} M_{\rm halo} v_{\rm c}^2$. That may be why simulations producing outflows with quite different mass-loading factors, like EAGLE and IllustrisTNG, can produce galaxy populations that are quite similar. 

However, outflows with similar power but different proportions of mass loading have different specific energies and therefore affect the CGM differently. EAGLE produces outflows with high specific energy, resulting in a more extended CGM and less recycling of gas through the central galaxy. IllustrisTNG produces outflows with less specific energy, resulting in a less extended CGM and more recycling, while greater mass loading compensates for the additional recycling.

We therefore suspect that specific energy may be the key to understanding why simulated galaxies having similar star-formation properties can have CGM characteristics that significantly differ. We also encourage other groups to perform similar specific-energy analyses on their own simulations, with more quantitative rigor than the sketch presented here.

\section{Summary}
\label{sec:Summary}

This paper has interpreted and discussed the implications of Paper I, which presented a new regulator model for galaxy evolution: the \textit{minimalist regulator model}. It prioritizes simplicity and generality over complexity and detail and focuses on how feedback heating and radiative cooling drive atmospheric expansion and contraction. 

Here is what Paper II adds to Paper I:
\begin{enumerate}

    \item Section \ref{sec:Equilibrium} describes the minimalist regulator model more concisely than Paper I does, outlining how the model makes it possible to define equilibrium states for galactic atmospheres. In those equilibrium states, the CGM's baryon fraction $f_{\rm CGM}$ and mean specific energy $\varepsilon_{\rm CGM}$ remain constant, along with the central galaxy's star formation rate $\dot{M}_*$, as long as the model's other parameters remain constant. Real galaxies may never settle into those equilibrium states because of fluctuations in cosmological accretion and supernova feedback. Nevertheless, the equilibrium states can be quite helpful for interpreting long-term trends.

    \item Observations that constrain the specific energy of galactic outflows may help distinguish \textit{preventative} feedback mediated by coupled outflows from \textit{ejective} feedback mediated by uncoupled outflows (\S \ref{sec:Ejective?}). We note that many observations probe only the denser parts of the outflows, but here we are particularly interested in the component carrying most of the energy, which is often diffuse and difficult to observe. If an outflow's specific energy $\varepsilon_{\rm fb}$ is much greater than $\varepsilon_{\rm acc} \approx 4 v_{\rm c}^2$, then its preventative effects on long-term galactic star formation are potentially much greater than its ejective effects, as long as the outflow eventually transfers its energy to the CGM. In contrast, ejective feedback needs to be fine-tuned such that $\varepsilon_{\rm fb} \approx \varepsilon_{\rm acc}$ in order to be maximally effective.
      
    \item Comparing the model's star-formation predictions with observations of the $f_*$--$M_{\rm halo}$ relation implies that the energy loading factor $\eta_E$ for feedback outflows depends on halo mass (\S \ref{sec:Predictions}). \citet{Carr_2023ApJ...949...21C} found that reconciling their model with observations required a mass dependence $\eta_E \propto M_{\rm halo}^{-0.6}$. Here, we find that the predictions of minimalist regulator models with $\eta_E \propto M_{\rm halo}^{-1/3}$ align with the data up to $M_{\rm halo} \approx 10^{12} \, M_\odot$. Beyond that borderline in halo mass, supernova feedback apparently cannot prevent contraction of the CGM and runaway star formation. Feedback powered by black hole accretion might then intervene to prevent a star formation crisis.

    \item Many cosmological numerical simulations of galaxy evolution employ parametric sub-grid models for supernova feedback, with parameter choices that optimize agreement with observations of galaxies but result in different outcomes for the CGM. The minimalist regulator model helps to explain why those outcomes are different (\S \ref{sec:Simulations}). It predicts that supernova feedback models resulting in outflows with greater specific energy and less mass loading should lead to more expansion of the CGM, smaller baryon fractions within $R_{\rm halo}$, and larger closure radii. 
    
\end{enumerate}

\begin{acknowledgements}

GMV acknowledges support from the NSF through grant AAG-2106575 and also benefited from the \textit{Turbulence in Astrophysical Environments} program, supported in part by grant NSF PHY-2309135 to the Kavli Institute for Theoretical Physics (KITP). GLB acknowledges support from the NSF (AST-2108470, ACCESS PHY2400043, AST-2307419), NASA TCAN award 80NSSC21K1053, the Simons Foundation (grant 822237) and the Simons Collaboration on Learning the Universe.
NASA provides support for VP through the NASA Hubble Fellowship grant HST-HF2-51489 awarded by the Space Telescope Science Institute, which is operated by the Association of Universities for Research in Astronomy, Inc., for NASA, under contract NAS5-26555. The Flatiron Institute is a division of the Simons Foundation. BDO's contribution was supported by Chandra Grant TM2-23004X. MD is grateful for partial support of this work from NASA award NASA-80NSSC22K0476.
\\

%\marknotes{Rachel -- Please add your acknowledgements here}
%rss it's in there -- i know it's a little strange, but my %acknowledgement is "The Flatiron Institute is a division %of the Simons Foundation"

\end{acknowledgements}

\bibliography{CGM}{}
\bibliographystyle{aasjournal}

\end{document}